\documentclass[pdflatex,sn-mathphys]{sn-jnl}

\usepackage{bookmark}
\usepackage{xcolor}
\usepackage{ulem}

\jyear{2021}%

\theoremstyle{thmstyleone}%
\theoremstyle{thmstyletwo}%

\theoremstyle{thmstylethree}%

\begin{document}

\title[Authentication of Underwater Assets]{Authentication of Underwater Assets}


\author*[1]{\fnm{Bálint Z.} \sur{Téglásy}}\email{balint.teglasy@ntnu.no}

\author[2]{\fnm{Emil} \sur{Wengle}}\email{emil.wengle@ntnu.no}

\author[2,3]{\fnm{John R.} \sur{Potter}}\email{john.r.potter@ntnu.no}

\author[4]{\fnm{Sokratis} \sur{Katsikas}}\email{sokratis.katsikas@ntnu.no}

\affil*[1]{\orgdiv{Department of Engineering Cybernetics}, \orgname{NTNU}, \orgaddress{\street{O. S. Bragstads Plass 2D}, \city{Trondheim}, \postcode{7034}, \country{Norway}}}

\affil[2]{\orgdiv{Department of Electronic Systems}, \orgname{NTNU}, \orgaddress{\street{O. S. Bragstads Plass}, \city{Trondheim}, \postcode{7034}, \country{Norway}}}

\affil[3]{\orgdiv{Centre for Geophysical Forecasting}, \orgname{NTNU}, \orgaddress{\street{O. S. Bragstads Plass}, \city{Trondheim}, \postcode{7034}, \country{Norway}}}

\affil[4]{\orgdiv{Department of Information Security and Communication Technology}, \orgname{NTNU}, \orgaddress{\street{Teknologivegen 22}, \city{Gjøvik}, \postcode{2815}, \country{Norway}}}


\abstract{Secure digital wireless communication underwater has become a key issue as maritime operations shift towards employing a heterogeneous mix of robotic assets and as the security of digital systems becomes challenged across all domains. At the same time, a proliferation of underwater signal coding and physical layer options are delivering greater bandwidth and flexibility, but mostly without the standards necessary for interoperability. We address here an essential requirement for security, namely a confirmation of asset identities also known as authentication. We propose, implement, verify and validate an authentication protocol based on the first digital underwater communications standard. Our scheme is applicable primarily to AUVs operating around offshore oil and gas facilities, but also to other underwater devices that may in the future have acoustic modems. It makes communication including command and control significantly more secure, and provides a foundation for the development of more sophisticated security mechanisms.}




\maketitle

\section{Introduction}\label{sec1}

Underwater (UW) environments are increasingly explored and developed for economic benefit, environmental stewardship and research interests. While workhorse-class Remotely-Operated Vehicles (ROV) will continue to play an important role in UW operations, due to requirements for substantial power and/or live video feed, their tethers can weigh several times the ROV itself, dramatically increasing power consumption to move them through the water, reducing maneuverability and creating entanglement and snagging issues \cite{RN811}. The proliferation of affordable light and agile Autonomous UW Vehicles (AUV) enabled by dramatic improvements in battery technology, cheap and ample processing and memory, developments in control theory and Artificial Intelligence (AI), etc., is empowering a disruptive technology change that is sweeping the field. 

Wireless UW Communications and Networking (WUCaN) is essential to support this new wave of autonomous systems. However, WUCaN is currently severely constrained compared to wireless communications in air, not only because of the formidable physical limitations, but also because few standards exist to support inter-operability. Currently the only open standard for UW wireless digital communications, a precursor to a fully-fledged WUCaN capability, is JANUS \cite{RN745}. Currently, WUCaN, if available at all, is generally conducted via unencrypted bitstreams without an authentication mechanism. In this paper we address and resolve this key shortfall.

We are essentially striving to create an Internet of Underwater Things (IoUT), by which we mean a Wide Area Network (WAN) of inter-operable UW devices. Just as the above-water Internet of Things (IoT) is based on radio links, we would also like a wireless solution. We look for potential authentication methods with an approach that, in principle, is agnostic to the physical layer, including radio frequency electromagnetic, free space optical and acoustic. Radio solutions are generally of very short range (O($10^0$)m) but have the benefit of potentially bridging the air-sea interface \cite{CHE2012317}. Free-space optical solutions have a larger, but still very limited, range of O($10^1$)m.  Both offer superior bandwidth compared to an acoustic physical layer, at the cost of very limited range. Only the acoustic physical layer has an accepted digital standard. Ultimately, we expect WUCaN systems to be intelligent, adaptive and physical-layer agnostic, but at this initial stage we begin with the most common physical layer, namely acoustics. We explore a baseline solution for civilian authentication requirements, develop a feasible method, and propose an attractive candidate for underwater assets using the JANUS protocol. It is intended primarily for use with AUVs, operating around offshore oil and gas facilities, to improve safety and productivity \cite{RN812}. The lack of authentication has always been a a primary concern in maritime communications, allowing countless false flag operations throughout history \cite{politakis2018modern}. As far as AUVs are concerned, it is easily imaginable that assets would be captured by adversaries due to the lack of secure communications \cite{SouthChinaAUV}, be it by knowing the location or even sending illegitimate command signals.

The remaining of the paper is structured as follows: In Section \ref{related work} we briefly review related work. In Section \ref{requirements} we specify the requirements that an authentication method for the IoUT should satisfy. In Section \ref{proposed} we present and discuss our proposal, including how it was implemented, verified, and validated. Finally, Section \ref{conclusion} summarizes our conclusions and outlines directions for future research.

\section{Related work} \label{related work}

Underlying physical layer technologies are advancing fast and international standardisation efforts are gaining traction, e.g. \cite{RN747}, which represents a bottom-up effort to achieve inter-operability driven by user necessity. WUCaN security threats are discussed by Yang et al. in \cite{RN809} while Peng et al. \cite{RN808} offer an encryption algorithm for UW use that is more energy efficient than previous solutions, although the block size of 64 bits poses questions of applicability in a standardised environment. Du et al. \cite{RN810} present a secure routing scheme for WUCaN, but the encryption method enabling their scheme is not defined in detail and it is not built on an existing physical protocol layer. Dini et al. propose a secure network discovery protocol for WUCaN in \cite{RN807}, where they primarily consider networks established between AUVs. The encryption method, the details of the physical protocol layer and the packet size (specific clear-text length) are not, however, developed. Petroccia et al. \cite{PetrocciaKobe2018} report network discovery and encryption with AES in Galois counter mode in the framework of their Cognitive Communications Architecture \cite{petroccia2018development}. This is a promising approach for interoperability as well, since JANUS is one of the physical layer protocols that the architecture is claimed to use. However, an authentication solution has not been described yet. In \cite{khalid2020node} and \cite{diamant2018cooperative}, communication security for underwater acoustic networks (UWANs) is addressed based on physical security, rather than point-to-point or sequential deterministic authentication. They note that UWAN packets are rarely encrypted, leaving the UWAN exposed to external attacks faking legitimate messages. This is essentially the problem we seek to address with cybersecurity methods. They propose a new algorithm for message authentication by observing that, due to the strong spatial dependency of the underwater acoustic channel, an attacker can attempt to mimic the channel associated with the legitimate transmitter only for a small set of receivers, typically just for a single one. Their scheme relies on trusted nodes that independently help a sink node in the authentication process.  For this to happen, we have to start with a set of trusted nodes. Then, for each incoming packet, the sink fuses beliefs evaluated by the trusted nodes to reach an authentication decision. These beliefs are based on estimated statistical channel parameters, chosen to be the most sensitive to the transmitter-receiver displacement. They have simulation results and at-sea experiments demonstrating the effectiveness of their approach. However, their approach relies on spatial dependencies and therefore on physical security; an attacker with an acoustic modem planted on or in the immediate vicinity of a trusted node is not defended against. Here, our method not relying on physical, but logical security in the form of a pre-shared secret provides a solution. An encrypted communication solution for JANUS, including packet formats for cargo length specification, has been suggested in \cite{RN803}. The encryption method is intentionally left to  the technology supplier or modem manufacturer using the JANUS standard, and the reception of more than a baseline packet is required to enable successful decryption. This solution, while promising, relies on a larger packet not being corrupted and an extension to the baseline JANUS standard. \cite{GhannadrezaiiJanus} assumes access to a hybrid system with radio communication, a public key system and AES encryption with a block size of 128 bits in the acoustic domain. With a slightly larger coverage of digital signature schemes, \cite{souza2013end} also assumes the presence of a network of base stations as an infrastructural pre-condition without getting into detail on how those base stations would be moored, powered or communicated with on a global scale. While public key systems undoubtedly have advantages for securing global systems for communication where the participating devices have no pre-shared keys, these rely on a likewise global public key infrastructure (PKI) to uphold the security properties promised by them. The communication requirements of a PKI would mean that ad hoc networking is not necessarily secure if the authorities in the infrastructure are not available e.g. through acoustic/radio gateways.

In Venilia \cite{Venilia} we see many of the same constraints being applied as in our proposal. A symmetric encryption scheme with an even smaller block size is used and epochs based on onboard time are harnessed to generate subkeys through a scheduler. However, there is no authentication protocol or other mechanism to ensure key renewal, such that the security property of forward secrecy \cite{boyd2021modern} is neglected. This is not acceptable in an environment where the scalability of mission duration or of the number of devices is needed. While the loss of confidential information such as keys is always unfortunate, it is catastrophic in the case of systems that only allow the use of a single key for all participants at all times. Venilia includes the routing data in the ciphertext as a sign that only one key can be used in one operations theatre. The risk of compromised keys through physical tampering of individual devices or any other cyber attack surface puts the whole fleet at risk, especially if the remaining payload of 8 bits is used for command and control as proposed. Nevertheless, we see the utility of Venilia in cases where many messages have to be sent back and forth including demands for checkbacks issued randomly, as would be the case for devices at lower levels of autonomy that need constant piloting. In these very limited cases, we concur that the non-determinism of Venilia guaranteed by initialisation vectors (IV) and epochs offers superior security.

To the best of our knowledge, no standardisable solutions for simple WUCaN authentication have been proposed in the literature. Accordingly, the purpose and contribution of this paper is to develop an attractive authentication method. To sum up, our proposed security barrier provides the flexibility to work as a local solution like Venilia, but also has key elements required for a more flexible and scalable solution without requiring infrastructure that would be unreasonable to assume.

\section{Requirements specification} \label{requirements}

\subsection{Choice of the Physical Layer}

Examples of the approximate bandwidth and range limitations of physical layer technologies are provided in Table \ref{physicallayers}. We are developing a system that is aware that there are different physical layers of interest, and provides defense in depth by authenticating with additional factors and bands as decreasing range allows.

\subsubsection{Electromagnetic}
Even though communication in the electromagnetic domain is severely restricted underwater, the possibility to use protocols such as the familiar 802.11b,g provides a tempting interface with enterprise systems, including the established authentication protocols on those systems (e.g. based on Kerberos \cite{RN817}, TACACS \cite{RN819} or RADIUS \cite{RN818}). This physical layer also offers a bridge to connect IoT with IoUT, a major issue in its own right.

\subsubsection{Free space optical}
Laser diodes with a 520 nm wavelength, modulated with Non-Return-to-Zero On-Off Keying (NRZ-OOK), have achieved a data rate of 500 Mbps with a bit error rate of 2.5 x $10^{-3}$ through clean freshwater in a laboratory \cite{RN784}. A blue laser (450 nm wavelength) optical modem is now commercially available that claims a robust data rate of 1 Mbps up to 15 m range in practical seawater applications with Ethernet compatibility.



\begin{table}[ht]
\begin{center}
\caption{Physical Layers for Underwater Communication}
\label{physicallayers}%
\begin{tabular}{@{}llll@{}}
\toprule
Modality & Reference  & Bandwidth & Range\\
\midrule
Electromagnetic&2,4 GHz WiFi \cite{RN749}& 11 Mbps& 15 cm$^{\mathrm{a}}$  \\
Free space optical& NRZ-OOK 520 nm \cite{RN784}& 500 Mbps& 100 m  \\
Acoustic&JANUS standard \cite{RN745}&80 bps& 10 km  \\
\botrule
\end{tabular}
\footnotetext{$^{\mathrm{a}}$\cite{RN749} indicates that packet loss rises steeply above 15 cm.}
\end{center}
\end{table}

\subsubsection{Acoustic} \label{background}
Useful UW acoustic communication frequencies span from O($10^0$)-O($10^6$)Hz, depending on the desired range and bandwidth considerations, but typically a modem in the 20-30 kHz range might offer O($10^0)$ kbps over a range of $\approx5$ km. There are many different physical layer protocols for digital acoustic communication, but they are all proprietary and therefore not interoperable, and also the extent to which academic inquiry is possible is limited. Furthermore, in cases where robustness is required, e.g. in noisy environments, the previously mentioned JANUS standard is as of 2021 still the fallback technology \cite{Mangi21}. JANUS was developed as a deliberately simple and robust physical layer protocol suited for initial contact, that could be used as a beacon, for discovery and for negotiation of mutually-available higher-performance communication modes, a function demonstrated in \cite{RN786}. As such, for lightweight authentication, the JANUS standard, with an 80 bps data rate (using the 11.520 kHz centre frequency specified for the first defined JANUS band), is very suitable. A complete communications system based on the JANUS physical and MAC layer protocols can be phrased in Open Systems Interconnect (OSI) terms as shown Table \ref{JANUSstack}. Whilst JANUS as a physical layer is comparatively simple, it does implement frequency-hopped binary shift keying to provide robustness in the face of multiple signal arrival paths. 

\begin{table}[ht]
\begin{center}
\caption{JANUS implemented in the OSI stack framework}
\label{JANUSstack}%
\begin{tabular}{@{}llll@{}}
\toprule
ISO OSI number & Protocol layer& Digital acoustic equivalent\\
\midrule
7& Application&   \\
6& Presentation& Implementation in \\
5& Session& non-standardised applications\\
4& Transport& (e.g. WetsApp)\\
3& Network& \\
\hline
2& Data link& partially covered by JANUS$^{\mathrm{a}}$\\
\hline
1& Physical& JANUS core specification\\
\botrule
\end{tabular}
\footnotetext{$^{\mathrm{a}}$JANUS includes the Medium Access Control (MAC) sublayer.}
\end{center}
\end{table}

When exploring the service support to be expected from the standard protocol stack, we begin by looking at the data link layer. JANUS includes a Cyclic Redundancy Check (CRC), but other functions of the data link layer such as flow control, acknowledgment, and error notification are absent. This means that all communications are unacknowledged and formally connectionless \cite{RN825}. The JANUS protocol also includes cross-layer features that may compromise strict adherence to the OSI layer architecture. As of writing, most of the OSI layers are implemented by non-standard user-defined applications. Although non-specified protocol layers facilitate the development of proprietary applications in a geographically and organisationally-segmented WUCaN market, they ultimately limit the inter-operability of networked and secured communication functions, unless they are co-ordinated with major stakeholders and become extensions of the standard. The absent protocol layers also mean that it is not possible to determine which packets arrived and were successfully decoded using only the baseline JANUS protocol. These challenges can be addressed by developing additional protocol elements, but for these to be useful, they must be simple and align with the JANUS philosophy of inclusivity, so that they are attractive to becoming intuitively adopted by the community. We account for this by designing the simplest possible protocol in this first iteration. This means using symmetric cryptography, as the distribution of public keys would impose an additional communication overhead and a public key infrastructure. It also implies using server-less protocols, because means of communication through centralized nodes and segmented networks are not likely to be available. To navigate the protocol stack and to have our packets be interpreted as part of the proposed protocol, Class IDs would need to be assigned.

\begin{table}[ht]
\caption{JANUS Bit Allocation in the Baseline Packet}
\label{tab:bits}
\centering
\begin{tabular}{|p{0.9cm}|p{1.6cm}|p{9.3cm}|}
\hline
\textbf{Bits} & \textbf{Descriptor}& \textbf{Comments} \\
\hline
1-4& Version& JANUS defined: unsigned 4 bit integer. Current version is 3. \\
\hline
5& Mobility flag& JANUS defined: Indicates nature of the transmitting platform.\\
\hline
6& Schedule flag& JANUS defined: If On (1), the first bit in the Application Data Block (ADB) indicates a cargo length. For our method, it is off.\\
\hline
7& Tx/Rx Flag& JANUS defined, Transmit/Receive capability: for our purposes, it needs to decode on both devices (1).\\
\hline
8& Forward capability& JANUS defined: Used for routing and Delay Tolerant Networking. For us,it should be 0=no. \\
\hline
9-16& Class User ID& JANUS defined: Allows 256 classes of users, mostly individual nations.\\
\hline
17-22& Application Type& Allows 64 different types of message per class user i.d. to be specified.\\
\hline
23-56& ADB& 34 bits of payload. Our proposal: 29 bit timestamp, 3 bit clock accuracy descriptor, 2 cleartext flags.\\
\hline
57-64& 8-bit Checksum&JANUS defined: 8-bit CRC run on the previous 56 bits with \(p(x) = x^8 + x^2 + x^1 + 1, init=0 \)\\
\hline
\end{tabular}
\end{table}

\subsection{Choice of an appropriate encryption algorithm for authentication}

Symmetric encryption methods are feasible if a copy of the cryptographic key can be shared, e.g. via WiFi, together with the synchronisation of clocks at some convenient opportunity when the assets are proximate in air, perhaps while batteries are being recharged or the systems are being otherwise prepared for deployment. Protocols based on a multi-step challenge-response and/or handshake are avoided, since short and variable channel coherence and asymmetric links are characteristic of the UW acoustic channel and an overly-demanding exchange could lead to very long or failed authentication processes. Therefore we develop our solution using only the JANUS baseline packet, whose bit allocation is shown in Table \ref{tab:bits}. This packet is 64 bits long, precluding the use of the Advanced Encryption Standard (AES) where the cipher block size is 128 bits.  This is a typical problem in WUCaN, where data rates are typically O($10^{-5}$) of those enjoyed in the GHz radio world, so that overheads of all types must be drastically reduced. The predecessor of AES, the Data Encryption Standard (DES), has a block size of 64 bits, and is still secure in the adapted version with enlarged key size called Triple DES \cite{RN804}. However, encrypting the full 64 bit packet would result in a lack of standardised bit allocation as formulated by JANUS. A major concern is accidental integrity loss in transmission; therefore the CRC must remain unencrypted. If we only want to encrypt the user-defined 34 bits in a JANUS packet ADB, the range of encryption methods available is reduced further. To fit the ADB, ciphers having a block size of at most 32 bits are considered. Informed by the list used in \cite{hatzivasilis2018review}, we conclude candidate ultra-lightweight ciphers as: RC5 \cite{RN805}, Speck \cite{RN827}, Katan32 \cite{RN828}, Hummingbird-2 \cite{RN806} and Skipjack32. We include the TUBCipher \cite{TUBCipher} designed specifically for Venilia for comparison. The characteristics of these algorithms are shown in Table \ref{tab:algorithms}. We want to maximise security within the constraints. Since the envisioned key exchange algorithm does not limit the key size, we prioritise those ciphers, within the block size bounds, with a large key (at least 128 bits). These are Hummingbird-2 and RC5. The Hummingbird-2 cipher has been developed with micro-controllers in mind. The simple RC5 code suggests that it might work better in software. Furthermore, the RC5 cipher requires no IV, whereas the Hummingbird-2 is like a stream cipher in this sense since it does. The communication requirements of synchronising an IV would put an additional burden on the complexity and reliability of the acoustic communication. Even if pre-shared, the IV would need to be updated synchronously on A and B. That is not feasible due to the high packet loss. Consequently we select the RC5 cipher.

\begin{sidewaystable}
\sidewaystablefn%
\begin{center}
\begin{minipage}{\textheight}
\caption{Encryption algorithms with block sizes $\leq$ 34 bits }\label{tab:algorithms}
\begin{tabular*}{\textheight}{@{\extracolsep{\fill}}lcccccc@{\extracolsep{\fill}}}
\toprule%
Cipher & RC5 & Skipjack & Speck & Katan32 & Hummingbird-2 & TUBCipher \\
\midrule
Cryptanalysis available? & Yes (64 bit$^{\mathrm{a}}$)& Yes (64 bit$^{\mathrm{a}}$) & Yes & Yes & Yes & No\\
    
    Minimum block size [bits] & 32 & 32 & 32 & 32 & 16 & 27\\
    
    Maximum key size [bits] & 2040 & 80 & 64 (32 bit$^{\mathrm{a}}$) & 80 & 128 & 256\\
    
    Needs IV? & No & No & No & No & Yes & Yes \\
    
    Software optimised & Yes & No & Yes & No & No & N/A \\
\botrule
\end{tabular*}
\footnotetext{$^{\mathrm{a}}$The ciphers marked with these properties have been formulated with different block sizes, but not all block size variants have been subject to peer-reviewed cryptanalysis or have the same key size. Therefore, the variant with the given block size has been evaluated.}
\end{minipage}
\end{center}
\end{sidewaystable}

\section{The proposed protocol} \label{proposed}

\subsection{Identification of Friend or Foe}
In the following we propose a mutual authentication solution, capable of identifying AUVs with pre-shared, long-term key \(K_1\). The following steps are required, grouped :

1. Device A sends a 64-bit baseline JANUS packet with a 29-bit timestamp \(T_A\), a 3-bit clock accuracy descriptor \(CD_A\) in the ciphertext and two 1-bit flags in the unencrypted payload, with the packet header and CRC. The year is assumed known, and the specifying the day as mod(Julian day,6) allows 29 bits to encode milliseconds. The three bits describing the on-board clock accuracy span the O($10^{-4}$) to O($10^{-12}$) drift rates. Some functionality, such as current and speed estimation, hinges upon the availability of high accuracy, low drift synchronization of the clocks on devices A and B, such as it is achievable with chip scale atomic clocks \cite{kebkal2017underwater}, while also allowing for more widespread quartz technologies. The remaining two bits should be used to specify: (i) If an answer is expected as a next step in the authentication protocol (SYN), (ii) if the packet being sent is sent as a response to acknowledge an earlier packet (ACK). These remaining two bits also give an indication of which key should be used as per Table \ref{tab5}.

2. Device B receives and decodes the packet. If the received timestamp is within bounds and not used in the current key lifetime, then B responds by sending its own timestamp and clock accuracy descriptor \(T_B, CD_B\)  encrypted with a pre-shared key \(K_1\) chosen according to the JANUS-defined cleartext header (bits 1-22 in Table 3) identifying a set including Device A. An application type of the class ID in the JANUS cleartext header also needs to indicate that the message is to be understood as one within our authentication framework. The packet including the returned clock signal of B should have its ACK bit set to 1.

3. If device A successfully receives and decodes the returned packet it can estimate the time of flight of the first outgoing packet by the difference \(T_B\)-\(T_A\) and the reciprocal time of flight for the second packet by the difference \(T_{A2}\)-\(T_B\) where \(T_{A2}\) is its own timestamp at the point of decoding the received response, adjusted by its own (known) decoding and decryption time delay. Given reasonable assumptions about the asymmetry in flight time due to currents (much smaller than the speed of sound in water) and possible mutual clock drift since the devices were synchronised during the key exchange, device A can determine if the received time stamp \(T_B\) is within expected margins and also estimate the inter-AUV distance based on a simple calculation using the speed of sound \cite{RN826}. This third step is optional, since the time window for response validity is already given by the time-of-flight at maximum assumed range and the inputs to the session key calculation (below) are already established in steps 1 and 2. However, since distance is the primary determinant of physical security and safety, the principal enabler of better communications than the initially assumed JANUS \cite{CapacityDistance}, and there is no appreciable overhead associated with this additional step, we strongly recommend it is performed.

\begin{figure} [ht]
\tikzset{every picture/.style={line width=0.75pt}} 
\begin{tikzpicture}[x=0.75pt,y=0.75pt,yscale=-1,xscale=1]
\path (0,170); 

\draw (147,48.5) node  {\includegraphics[width=207pt,height=69.75pt]{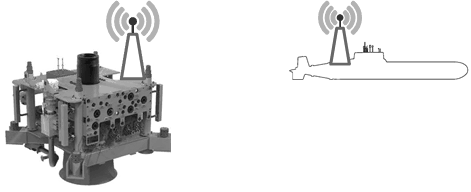}};
\draw    (145,106) -- (173.12,116.31) ;
\draw [shift={(175,117)}, rotate = 200.14] [color={rgb, 255:red, 0; green, 0; blue, 0 }  ][line width=0.75]    (10.93,-3.29) .. controls (6.95,-1.4) and (3.31,-0.3) .. (0,0) .. controls (3.31,0.3) and (6.95,1.4) .. (10.93,3.29)   ;
\draw    (145,1) -- (145,260) ;
\draw    (175,1) -- (175,260) ;
\draw    (175,167) -- (148.26,181.54) ;
\draw [shift={(146.5,182.5)}, rotate = 331.46000000000004] [color={rgb, 255:red, 0; green, 0; blue, 0 }  ][line width=0.75]    (10.93,-3.29) .. controls (6.95,-1.4) and (3.31,-0.3) .. (0,0) .. controls (3.31,0.3) and (6.95,1.4) .. (10.93,3.29)   ;

\draw (0,98) node [anchor=north west][inner sep=0.75pt]   [align=left] {{\small {\fontfamily{ptm}\selectfont 1. Device A sends  }}\\{\fontfamily{ptm}\selectfont{\small \{T}\textsubscript{A}, CD\textsubscript{A}\}\textsubscript{K1}}};
\draw (177,120) node [anchor=north west][inner sep=0.75pt]   [align=left] {{\fontfamily{ptm}\selectfont {\small 2. Device B within range  }}\\{\fontfamily{ptm}\selectfont {\small decrypts T\textsubscript{A}, CD\textsubscript{A}}}\\{\fontfamily{ptm}\selectfont {\small answers with}}\\{\small {\fontfamily{ptm}\selectfont \{T}\textsubscript{B}, CD\textsubscript{B}{\fontfamily{ptm}\selectfont \}}\textsubscript{K1}}};
\draw (2,177) node [anchor=north west][inner sep=0.75pt]   [align=left] {{\fontfamily{ptm}\selectfont {\small 3. Device A decrypts  }}\\{\fontfamily{ptm}\selectfont{\small
\{T\textsubscript{B}, CD\textsubscript{B}\}\textsubscript{K1},}}\\{\fontfamily{ptm}\selectfont {\small catches timing errors.}}\\{\fontfamily{ptm}\selectfont {\small All devices calculate \(K\textsubscript{AB}\)}}};
\end{tikzpicture}
\caption{An illustration of the authentication challenge \({\{ T_{A}, CD_A \}}_{K_1}\) and the response \({\{ T_{B}, CD_B \}}_{K_1}\).}
\label{fig:method}
\end{figure}
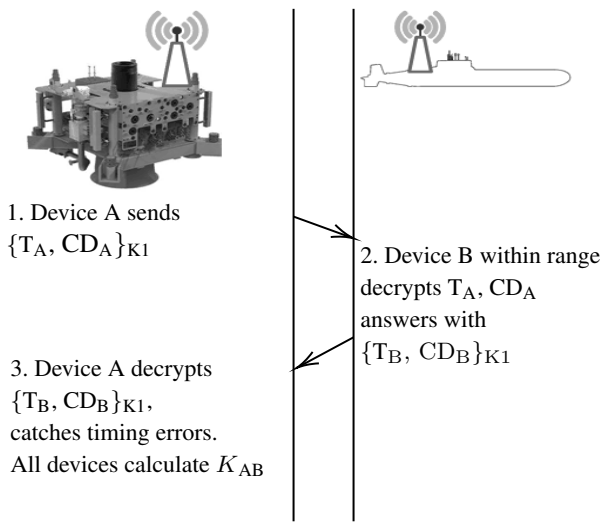

The proposed method is illustrated in Figure \ref{fig:method}, where an AUV and a subsea valve assembly typical for oil and gas production are depicted as communication partners.

\subsection{Calculating a Session Key}

As we have not yet negotiated a shared secret that can be used as a session key in following communications, our method so far might not qualify as full feature authentication. Instead, the main functionality provided up to this point is the positive identification of friends.

If the derivation of a common secret is desired, both of the devices can calculate that now based on the timestamp they received from the other device and the one they have transmitted last. Note that other devices C, D, etc. within the reception zone and in possession of the long-term key, therefore declared friendly in our security model, will also be able to derive the session key. Due to the uncertainty resulting from high packet loss rates, device A is in a better position to start using the session key as it has in step 3 received an ACK-flagged confirmation that its timestamp and clock descriptor \(T_A, CD_A\) are available to some friendly device B in possession of the long-term key \(K_1\). The common secret would be the pair of securely exchanged payloads \(MMSI_A, T_A, CD_A\) and \(MMSI_B, T_B, CD_B\). The fresh session key only available to friendly devices is \(K_{AB} = f(T_A, CD_A, T_B, CD_B, K_1)\), where \(f\) is a combining function that doesn't allow finding \(f(.,.,K)\) without knowledge of the long-term, pre-shared key \(K_1\) \cite{boyd1995towards}. For \(f\), we propose:

\begin{itemize}
\item concatenating \(MMSI_A, T_A, CD_A\) and \(MMSI_B, T_B, CD_B\),
\item bit padding the resulting 124 bits to 512 bits by appending one bit as 1 and 387 bits as zeroes (1000...0),
\item Apply the 128-bit block size version of the RC5 cipher in CBC (Cipher Block Chaining) mode with starting variable fixed to 0, as defined in ISO/IEC 10116
\item truncating the resulting ciphertext to the first 256 bits
\end{itemize}
 
By doing so, we have established forward secrecy: the communications under \(K_{AB}\) will remain secure even when the long-term key \(K_n\) is compromised, provided that \(T_A, CD_A\) and \(T_B, CD_B\) are not simultaneously compromised. Note that this third step to establish a session key could be used also when \(T_A, CD_A\) and \(T_B, CD_B\) have been exchanged in cleartext: this just removes the dependency on \(K_1\) along with the trust that it brings, but this might be necessary if no \(K_1\) is available and instead physical layer security is deemed to be sufficient.

Device A could send \(T_{A2}\) as a confirmation under \(K_{AB}\) to device B. In the event that B has the less accurate clock, this provides an unambiguous estimate of the differential clock offset and water current velocity projected onto the vector joining the two devices. Additional functionality to synchronize clocks and/or estimate current and/or vehicle speeds can be based upon the exchanged clock accuracies. If one of the authenticated devices has a better clock accuracy than the other, the one with the less accurate clock should synchronize its own by taking the time stamp of the other device as its own (after adding half of the round-trip time). If this is done correctly, the chance of future successful authentications among the same devices increases. In a model with a variety of devices running different clocks and authenticating with each other at different intervals, this would help ensure that the clocks stay synchronized.

\subsection{Exchanging Unique Identifiers, Renewing Long-Term Keys and Further Ranging}

In the first three steps, we have provided three of four desired properties of a key agreement protocol. These properties were defined in \cite{boyd1995towards} as follows:
\begin{enumerate}
\item Both participants possess \(K_{AB}\) which they can verify is new.
\item It is infeasible to find \(K_{AB}\) by eavesdropping on the protocol, even if the protocol is repeated many times.
\item Both participants have equal input into the equation that defines \(K_{AB}\).
\item Both participants know the identity of the other party who may possess \(K_{AB}\).
\end{enumerate}

Regarding the fourth property, our solution so far is not necessarily satisfactory. A pre-shared table of \(K_{n}\) with corresponding identities might be used, where the encrypted payload is decrypted with every \(K_{n}\) in the table and the identity is assigned according to the table if one of the timestamps yields a successful authentication. This solution is sub-optimal for two reasons: (1) the false positive rate for adversaries trying to guess the key is increasing with every new \(K_{n+1}\) in the table. Although the unusually large key size alleviates these concerns for \(n<1000\), it is still not the scalable solution we are looking for when we aspire for interoperability. (2) the decryption attempts take time and energy. The time component adds complexity to the error-catching based on timestamps.

Assuming that a Class User ID (bits 9-16 in Table IV) can be reserved for our authentication solution, using the cleartext application type allows a receiver to look up one of 64 keys to be used for decrypting messages. The lookup table used for this purpose should have a unique identifier for each device as a primary key. For this purpose we propose a version of the Maritime Mobile Service Identity (MMSI) to be pre-assigned to all marine assets capable of wireless communication. We believe this to be the trend regardless of our underwater communication efforts \cite{ferreiraASVreg}. The AIS (Automatic Identification System, for tracking ships) builds on MMSI, therefore the fusing of surface and UW assets can be achieved easily if both carry the same individual identifiers. The 9 decimal digits of the MMSI are converted into 30 bits, as such they conveniently fit the 32-bit payload. We have therefore found a way to secure the exchange of unique identifiers. While AIS uses its own physical layer protocol based on ISO/IEC 13239:2002 and has its own proposals for securing it, e.g., \cite{goudosis2020secure}, the establishment of an underwater AIS seems feasible if the MMSI can be relayed through an acoustic/radio gateway.

A and B can thus securely negotiate much higher bandwidth and/or lower packet loss physical layers, such as those described in \cite{chitre2008recent} or \cite{kilfoyle2000state}.

At the end of the protocol, it is prudent to delete \(T_A, CD_A\) and \(T_B, CD_B\) from memory after deriving \(K_{AB}\), so that the capture of A or B would not enable an adversary to derive \(K_{AB}\) and decrypt previously recorded messages with it. The derived session keys should instead be stored and looked up in a table where the JANUS cleartext header determines the session key to be used. For further communications with the session key, the cleartext SYN and ACK flags should both be set to 0 and 1. Since the introduction of the session key it is not straightforward which key, if any, the devices should use to try to decrypt communications. The keys could all be tried and the cleartext fitted to expectations, but that would require more than necessary computational power and complexity. The following table provides clarification:

\begin{table}[ht]
\caption{Flag and key use for the protocol messages}
\label{Flags}
\centering
\begin{tabular}{|c|c|c|c|}
\hline
\textbf{Message Number} & \textbf{SYN} & \textbf{ACK} & \textbf{Key to be used}\\
\hline
1.& 1& 0& \(K_n\)\\
\hline
2.& 1& 1& \(K_n\)\\
\hline
3. and following& 0& 1& \(K_{AB}\)\\
\hline
Wide-area transmission when required& 0& 0& according to MMSI\\
\hline
\end{tabular}
\label{flags}
\end{table}

However, since our session key is only 64 bits long and we used timestamps to derive it, resistance against brute-force attacks is not necessarily ensured in the long term. If secure communication between A and B is desired beyond 10000 packets, \(K_{AB}\) can be used as a key wrapper under which a longer key \(K_{2}\) is communicated between A and B. This could be the case if continuous data transmission is desired. The device making up \(K_{2}\), for example by randomly generating it, would assume the role A. If this long-term key is a new 2040 bit long-term key, it could be transmitted with a cargo length specified in less than a minute. A series of baseline packets would be possible, but that would waste time and therefore bandwidth due to the repeated need to encode identical headers. Instead, the schedule flag located at the sixth bit of the JANUS header should be set to 1 for this purpose. The 8 remaining bits in the encrypted payload of the last packet should be used to detect adversarial modifications of the long-term key with reasonable probability. This can be achieved by an 8-bit CRC \(p(x) = x^8+x^2+x^1+1\), initialised to 0 (as specified by the JANUS for cleartext use as well) calculated over the cleartext, and including that CRC in the ciphertext in addition to the unchanged CRC of the baseline packet. Having the CRC in the ciphertext will protect against adversarial as well as accidental modifications of the packet. The receiving device could confirm correct (as per encrypted and cleartext CRCs) reception of the \(K_{2}\) in one baseline packet with the CRC this time being encrypted under the new \(K_{2}\).

\begin{figure} [ht]
\tikzset{every picture/.style={line width=0.75pt}} 

\begin{tikzpicture}[x=0.75pt,y=0.75pt,yscale=-1,xscale=1]
\path (0,170); 

\draw (147,48.5) node  {\includegraphics[width=207pt,height=69.75pt]{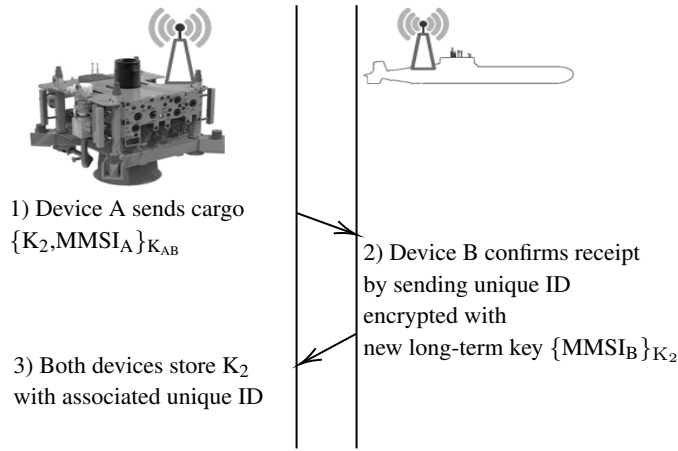}};
\draw    (145,106) -- (173.12,116.31) ;
\draw [shift={(175,117)}, rotate = 200.14] [color={rgb, 255:red, 0; green, 0; blue, 0 }  ][line width=0.75]    (10.93,-3.29) .. controls (6.95,-1.4) and (3.31,-0.3) .. (0,0) .. controls (3.31,0.3) and (6.95,1.4) .. (10.93,3.29)   ;
\draw    (145,1) -- (145,225) ;
\draw    (175,1) -- (175,225) ;
\draw    (175,167) -- (148.26,181.54) ;
\draw [shift={(146.5,182.5)}, rotate = 331.46000000000004] [color={rgb, 255:red, 0; green, 0; blue, 0 }  ][line width=0.75]    (10.93,-3.29) .. controls (6.95,-1.4) and (3.31,-0.3) .. (0,0) .. controls (3.31,0.3) and (6.95,1.4) .. (10.93,3.29)   ;

\draw (0,98) node [anchor=north west][inner sep=0.75pt]   [align=left] {{\small {\fontfamily{ptm}\selectfont 1) Device A sends cargo  }}\\{\fontfamily{ptm}\selectfont {\small \{K\textsubscript{2},MMSI\textsubscript{A}\}\textsubscript{K\textsubscript{AB}}}}};
\draw (177,120) node [anchor=north west][inner sep=0.75pt]   [align=left] {{\fontfamily{ptm}\selectfont {\small 2) Device B confirms receipt  }}\\{\fontfamily{ptm}\selectfont {\small by sending unique ID}}\\{\fontfamily{ptm}\selectfont {\small encrypted with}}\\{\small {\fontfamily{ptm}\selectfont new long-term key \{MMSI\textsubscript{B}}{\fontfamily{ptm}\selectfont \}}\textsubscript{K\textsubscript{2}}}};
\draw (2,177) node [anchor=north west][inner sep=0.75pt]   [align=left] {{\fontfamily{ptm}\selectfont {\small 3) Both devices store K\textsubscript{2}  }}\\{\fontfamily{ptm}\selectfont {\small with associated unique ID}}};
\end{tikzpicture}
\caption{An illustration of the option to renew a long-term key by wrapping it in the session key.}
\label{fig:longkey}
\end{figure}


\subsection{Unicast secure communication}

Once keys have been generated and stored along with unique identifiers of the devices, another application would be secure unicast communication. The unicast communication concept would enable the hardening of underwater communication security according to general cyber \textit{need to know} principles. This type of communication would necessitate the unique identifier of the sender to be sent in cleartext so that the corresponding key can be looked up by the recipient. A packet ensuring secure unicast would need to have the following pre-conditions:
\begin{itemize}
\item An application type is standardized in the JANUS header that identifies this unicast mode
\item Previous steps of our method for deriving a bilaterally shared session key \(K_{AB}\) have been successful
\item A cargo specification in the JANUS header, because the MMSI as a suitable unique identifier would already take 30 of the 34 bits in the ADB
\item A lookup table of MMSI and session key(s) on the recipient device.
\end{itemize}
This would allow packets of the format \(MMSI_B,{\{ payload \}}_{K_{AB}},HMAC\) to be transmitted securely to device B through a wide area network using the MMSI as an address. The cleartext inclusion of the MMSI is deemed necessary for inter-networking efforts, where devices who received the packet but were not the addressee may choose to re-transmit. The HMAC (keyed-hash message authentication code) should be calculated over the entirety of the packet including the JANUS header using the session key. This would authenticate the information there, most importantly the Class ID that determines which applications are to be used in interpreting the packet. Because of the bandwidth limitations, the HMAC also serves the purpose of a compressed sender designation: the recipient tries to decrypt the packet with all the keys found in its onboard database of MMSI and session key pairs. The session key that can verify the HMAC as authentic will indicate the correct MMSI of the sender in the lookup table.
When considering AUVs, the payload above could be a command and control signal, ideally compressed according to a pre-shared lookup table.

\begin{table}[ht]
\caption{JANUS Bit Allocation in the Unicast Secure Packet}
\centering
\begin{tabular}{|p{0.9cm}|p{1.6cm}|p{9.3cm}|}
\hline
\textbf{Bits} & \textbf{Descriptor}& \textbf{Comments} \\
\hline
1-22& Baseline header& JANUS defined as per above \\
\hline
23-30& Cargo length specification& JANUS defined: reserves the channel, in this case with i=60 for another second.\\
\hline
31-54& ADB& 24 bits of routing data from the MMSI range for autonomous systems (to be assigned)\\
\hline
55-56& Syn/Ack Flags& Aids the treatment of the packet as per Table \ref{Flags}\\
\hline
57-64& 8-bit Checksum&JANUS defined: 8-bit CRC run on the previous 56 bits with \(p(x) = x^8 + x^2 + x^1 + 1, init=0 \)\\
\hline
65-128& Encrypted Payload Cargo& 64 bits can be encrypted with RC-5 \\
\hline
129-137& HMAC& Calculated over the last 128 bits, it allows the recipient to authenticity of the message.\\
\hline
138-146& 8-bit Checksum& 8-bit CRC run on the previous 56 bits with \(p(x) = x^8 + x^2 + x^1 + 1, init=0 \)\\
\hline
\end{tabular}
\end{table}

\section{Results and Discussion}

\subsection{Mitigation of selected attacks}

As usual with authentication methods, assets participating in our protocol are classified into friendly (well-meaning) and malicious (adversarial) ones according to their ability to prove their identity. Methods to prove a false identity are considered attacks on the authentication method.

Due to our proposed design choice of using only the ADB, we note that the encrypted message can be changed without knowledge of the encryption key along with the cyclic redundancy check. This would allow adversarial submissions of valid JANUS packets which would fail to authenticate. Our proposed mitigation is to eliminate the possibility of attacks based on repeated submissions: error-catching should be implemented for valid packets with an already used timestamp in the decrypted ADB.

Protection against replay of earlier captured messages is achieved by validating the decrypted pongs against a time stamp. If the decrypted device B time stamp does not provide nearly-symmetric packet travel time estimates (allowing for currents and modelled mutual clock drift statistics and corrected for encryption and decryption processing delays), a failed authentication notification results. It is of course possible for an adversary to derive the cryptographic key used for authentication after observing and logging many authentications with that key \cite{RN821}, but our application is not likely to provide sufficient examples to enable this breach.

Challenge intervals should be informed by the expected maximum approach speeds. E.g. a challenge being sent out every 5 minutes would ensure that an AUV with a maximum speed of 3 m/s gets interrogated within a kilometer of entering the reception range. When rolling over the 6-day interval covered by the 29-bit timestamp, device A is advised to offset its challenge by 30 seconds to mitigate an attack where a ciphertext recorded 6 days ago is replayed. 30 seconds are deemed enough for the signal to be beyond range. These assumptions would result in the necessity to issue new keys every 60 days.

Denial of service (DoS) is possible through repeated re-transmission of earlier messages as well as the modification of the ciphertexts and the corresponding CRC. However, denial of service would also be possible without the proposed security method by making noise. This is the case for all wireless communications, but more so for acoustic communication. Therefore we assign the DoS challenges to the physical layer realm and do not provide design features to avoid them in this paper.

We believe that cybersecurity measures suggested for standardization today should also be vetted for resistance to quantum computers. As our scheme is based on symmetric cryptography, it is somewhat resistant. Furthermore, the unusually large key size gives us sufficient certainty that quantum-enabled algorithms like that of Grover \cite{bernstein2010grover} will not compromise our method.

\subsection{Ranging functionality and its possible ramifications}

Regarding the assumptions made in step 3 of the identification of friend or foe, we assume that over the timescale of the exchange, the primary eigenpath is reciprocal between A an B and does not change appreciably. This is likely the case if the path is not interacting with the sea surface because A and B are at depths exceeding 100 meters. Whilst the symmetry of the acoustic channel between A and B isn't necessarily given, there wouldn't be a successful exchange of data for a challenge and a response in such cases anyways. It might be possible to estimate such asymmetric channels for the use of coherent physical layers, but this wouldn't fit our requirement of using just one JANUS baseline packet each way. It is also likely that the additional complexity introduced with the use of such more advanced physical layers would come at the cost of lost interoperability.

By sending out the authentication challenge in cleartext, device A could give away its clock signal, and with it its location. This expression of trust is not advised in an adversarial environment, because it could enable cyber attacks based on the provoked exhaustion of the ciphertext space by an intruder masquerading as A, or physical attacks based on the necessary response from B confirming its presence. Nevertheless, our present protocol can be modified in line with civilian transponder interrogation such as Mode S in air traffic \cite{trim1990mode}, where collision avoidance and the avoidance of over-interrogation are priorities. Depending on the use case, this might be a proportionate measure to maintain the interoperability of JANUS across organisational borders while providing accountability. In \cite{ferreira2018increasing}  operational safety is seen to increase by sending location and heading data in addition to the MMSI. The two clock drifts do not impact the range estimate from our protocol, with vehicle motion and water currents contributing only second-order errors, as we show in Appendix \ref{Clock}.

\subsection{Applicability for different underwater assets}

Authentication services create a foundation for an IoUT.  Our authentication method can be generalised to a wider range of subsea assets than AUVs; all devices with an acoustic modem would profit from an inter-operable authentication method. Nevertheless, the requirement for secured wireless communication imposes constraints, e.g. cryptographic keys must be securely distributed. The mobility of AUVs makes key distribution through UW WiFi or short-range directive optical communication (which is remarkably secure to interception) easier, and more feasible on a regular basis, than between fixed assets. If new keys cannot be exchanged regularly and if there are many authentication attempts, security might be compromised. For a heterogeneous system that includes both fixed and mobile assets, keys might be regularly exchanged and clocks synchronised by an AUV mule activity, in which an AUV would visit all other assets in the system to perform key exchange and synchronisation by very short-range directive optical communication, which presents a much more challenging task to break into compared to omnidirectional acoustic signalling. In the case of asset classes that can't or don't want to exchange new keys, key derivation should be considered. This could be done using any one-way function, if parts of the initial long-term key and the calendar year and week are inputs to that function.

However, the applicability of the authentication method presented here for use in authenticating more than two devices simultaneously -- meaning more than a bilateral relation -- is limited. If in addition to devices A and B, a friendly device C is within hearing distance, it could derive wrong MMSI/session key pairs. This shortcoming is not relevant for most underwater economic ecosystems today, but it could be in a future where several previously unidentified friendly devices answer the same call. While mitigation is possible by setting the time window validity lower, we seek a solution that is more scalable. This problem is among those that we intend to address in our future research.

\subsection{Verification and Validation}

We implemented the proposed authentication protocol and tested it in air, in a small water tank, and in seawater in an outdoor harbour environment in the Trondheim fjord.\footnote{In addition to these physical tests, in silico testing was performed using the Network Simulator 3 Underwater Acoustic Network (NS3 UAN) library. Electronic supplementary material has been made available to JMST to aid reproduction of all claimed results so far as well as further verification and validation.} (B2) Two Subnero Research Edition modems were used in these tests. We had to write our own implementation of RC5 in Java for the agent to call, as the UnetStack Software-Defined Open-Architecture Modem Audio Driver is written in Java/Groovy. (B3) The validation tests successfully demonstrated the authentication protocol by deriving the same session key on the two devices, and put an upper bound of 3 seconds on all communication overhead resulting from the implemented security countermeasure. This overhead could likely be decreased by optimizing the hardware and software.


Our requirements, the specifications we derived from them and fulfilled with our solution can be summed up as follows:

\begin{table}[htbp]
\caption{Requirements and Specifications in the Authentication of Underwater Assets}
\centering
\begin{tabular}{|c|c|}
\hline
\textbf{Requirement} & \textbf{Specification}\\
\hline
Minimized number of packets& 2 packets sufficient for friend ID\\

Fits JANUS baseline ADB& 34 bits\\

Range at least 10 km& 10+ km for 11 kHz acoustics\\

Key size at least 256 bits& 2040 bits\\

Allows autonomous bilateral ranging& Through redundant timestamps\\

Run-time demonstrated & 3 seconds\\

\hline
\end{tabular}
\label{tab5}
\end{table}

Before being put to use in industry, a new technology or procedure needs to be extensively verified and validated in several steps.
By verification we mean testing that the technology or process meets requirements. At least three verification steps are recommended to be performed \textit{in silico}:

Firstly, crypto-analysis of the RC5 variant proposed to exchange the first two messages\footnote{In the notation RC5-w/r/b, where w=word size in bits, r=number of rounds, b=number of 8-bit bytes in the key, the variant satisfying our size restrictions and maximising security beyond that which would be known as RC5-16/255/255.} should be sought. The small block size might be exploitable, whereas the larger than usual key size and the high number of rounds could compensate for that. Based on crypto-analytic results, the number of rounds might be decreased if reasonable security can be achieved despite the minimal block size.

Secondly, the encryption, coding, decoding and decryption times with the hardware and software available on the UW assets should be characterised, together with the mutual clock drift statistics.

Thirdly, transmission technologies contributing to physical layer security such as predictive beamforming should be integrated in the UW assets \cite{RN1230}. This kind of development would be greatly accelerated by using state of the art digital twins \cite{RN1231} \cite{RN1232} including representative propagation modelling and adequate computational power and memory.

After the verification phase has been initiated, and partially overlapping with it to provide iteration opportunities, operational strategies and technologies should be clarified through validation. The validation stage should include testing unforeseen difficulties with AUVs, in addition to testing already done with modems suspended from the surface. In real WUCaN use cases that impact economic and usability aspects this could involve the tie-in of orthogonal security cross-checks, such as sonar object recognition, so that more rigorous authentication challenges can be directed at unidentified assets in proximity.

After the method has been rolled out as a pilot project, corporate security audits could use documentation from the verification and validation phase to inform their judgement of underwater communications. This could include penetration testing through partially UW red team exercises, with the validation goal to prove inability to obtain friendly identification without initial knowledge of the key.

\subsection{Authentication and Safety}

The importance of communication using the JANUS standard for operational safety is discussed in \cite{ferreira2018increasing}, where it is being implicitly assumed that there are only honest underwater assets. Authentication as a foundational requirement for security can help ensure that those operational safety goals are upheld in environments without total trust. Collaborative safety mechanisms have cybersecurity as a cornerstone technological necessity \cite{IECSafetyFuture}.

It is conceivable that AUVs will be credited as a safety barrier for mitigating oil and gas blowouts, similar to how ROVs worked to contain the Deepwater Horizon spill. Many AUVs work concurrently in mitigative scenarios. Valves operated by AUVs in a safety-critical setting can include all-electric valves on the Christmas Trees permanently located subsea \cite{RN830}. Such solenoid valves could need to be operated by AUVs, e.g to connect emergency power or apply the torque from batteries or motors on an AUV. If the Christmas Tree has an acoustic modem and a wired connection to a control room, operators can use the proposed authentication method to ensure that an AUV with the right key and working acoustic communication is approaching. The authorisation following authentication should be considered an essential service (as in the IEC/ISO 62443 series of international standards, henceforth 62443) for safety-critical resources. While describing the fundamentals of our authentication method, we have employed a pair of pre-shared keys K. In the framework of a more sophisticated access control scheme, an almost arbitrary variety of long-term keys \(K_{1}\) to \(K_{1,26e+614}\) can be issued to different roles or organisations. If this option is used, the long-term keys should be tried for every authentication attempt where no contextual information is available on the claimed unique identity. This will serve to reduce the false negatives due to the use of different keys. Due to the large key size, the false positives will not rise significantly by doing so. The simple approach outlined here may thus be extended not only to assets beyond AUVs but also to more complex and secure nested systems. The key size offers also the opportunity to establish national and global systems, if key management services are provided by maritime authorities. The details of such a key management scheme shall be described in further research, in the meantime it suffices to say that the number of operating organisations is virtually unlimited.



\subsection{Compliance with standards}
As has been amply demonstrated by IoT developments above water, there is potentially great cost to users who do not establish sufficient communication security and we can expect the same to be true for the IoUT. The  62443 imposes compliance specifications on the security in industrial communication networks. The scope of 62443-1-1, among others, specifically includes: (i)	oil and gas production operations as defined by functionality in chapter 1.2, (ii) activities necessary for predictable operation of the process in chapter 1.4, and (iii) assets needed for disaster recovery according to asset-based criteria in chapter 1.5. Based on these scoping criteria, it can be argued that AUVs used for inspections or disaster response are within the purview of 62443. Identification and Authentication Control (IAC) is the Fundamental Requirement 1 in 62443, and IAC influences the security levels (SL) assigned in 62443. This means that compliance with 62443 cannot be achieved as long as there is no authentication in all of the industrial automation and control system components. While UW devices are not yet networked, there is a strong incentive to do so, and if AUVs are networked without authentication, compliance with 62443 will not be possible.

The choice of entity authentication protocols is treated by the ISO/IEC 9798 family of standards, where part 2 concerns those methods using symmetric encryption algorithms. Our proposal is a refinement of the two-pass mutual authentication protocol described in chapter 7.3.2 of that standard, where we did not include a unique identifier of the recipient within the same encrypted package as the timestamp. This is a design option left open by the standard, as it can be also read in the clarification of the relevant standard section provided in \cite{boyd2003protocols}.

The part of step 3 of our proposal that establishes a session key uses a one-way function established in line with MAC Algorithm 1 described in the ISO 9797-1:2011 standard on Message Authentication Codes using a block cipher. It could be completely and unambiguously defined by the selection of the RC5-64/255/255 block cipher algorithm, padding method 2, and the length of 64 bits.

If, for whatever reason, Venilia \cite{Venilia} is to be used for encrypted communications, the authentication method we describe for deriving a session key might still be useful as an add-on. The 256-bit keys that the Tiny Underwater Block Cipher uses can be derived with our current proposal. This will then provide the forward security property highly recommended for a scalable solution and allow distance to be introduced as a risk metric, e.g. for further use in ensuring physical security or collision avoidance.

\section{Conclusion} \label{conclusion}

We have in this paper presented a draft protocol based on the first digital UW communications standard, JANUS. We believe that the initial idea behind JANUS as a ‘first contact’ handshake protocol is made significantly more secure by applying our protocol. Two friendly devices could be confirmed as such before deriving a session key under which they could securely negotiate another, hopefully higher bandwidth physical layer for further communications. The physical security elements of the newly negotiated physical layer would therefore remain confidential. 

While the timestamps we propose to be sent for authentication purposes enable ranging through a Time-of-Flight principle, we have not yet conducted tests to determine how accurately this ranging can be performed in practice. Depending on the accuracy of range authentication, actions of different criticality could be authorized. Factors influencing this accuracy will include, but are not limited to: modulation speed of the individual data packets, tick period length of the real time systems used to decrypt and encrypt, and variations in signal speed. Authorization should be defined on the basis of our authentication method to complete an access control framework for AUVs. In the absence of a docking station providing underwater WiFi connection to an enterprise-level authentication solution, AUVs could still derive long-term keys for encrypted communication. One potential solution could be the use of a shared medium or of Physically Unclonable Functions (PUFs) as a source of additional keys \cite{RN780,RN781}. This development can be observed in above water IoT radio frequency applications (Wi-Fi) that are also physically exposed \cite{ZENGER2016authN}, resource- and power-constrained \cite{Zenger2014}, and mobile \cite{staat2020intelligent}. By deriving keys from sonar signatures instead of radar, and acoustic instead of radio channel state information, similar security \cite{guillaume2015} could be provided underwater. We intend to pursue this line of research in the future.

\backmatter

\bmhead{Acknowledgments}

This paper has been written as part of the collaboration with IFE (Bjørn Axel Gran). The research was motivated by Vidar Hepsø as part of the BRU21 Research and Innovation Program on Digital and Automation Solutions for the Oil and  Gas Industry (www.ntnu.edu/bru21), and received funds from the Research Council of Norway, project number 302435, \textit{Autonomous Underwater Fleets: from AUVs to AUFs through adaptive communication and cooperation schemes}.

\begin{appendices}

\section{Establishing the theoretical accuracy of the ranging functionality}\label{Clock}

Suppose device A transmits an interrogating ping at T\textsubscript{0}. Device A has a clock offset \begin{math}\delta \textsubscript{A}\end{math}, which we will assume is constant over the authentication protocol execution.

If the distance between A and B is $d$, the speed of sound is $c$ and the current velocity along the line joining A and B is $v$, then B receives the ping at time
\[ T\textsubscript{1} = T\textsubscript{0} +\frac{d}{c+v}\]
Ignoring the motions of A and B (which can be included in v) and the processing times  (which, if known, can be subtracted out) then the ping timestamp $T_{A0}$ is given by
\[T\textsubscript{A0} = \delta \textsubscript{A}\]

Device B responds with
\[T\textsubscript{B1} = T\textsubscript{1} + \delta \textsubscript{B}\]
and device A receives this 'pong' at time $T_2$ given by
\[T\textsubscript{2} = T\textsubscript{1} +  \frac{d}{c-v}\]
so device A now knows
\[T\textsubscript{B1}-T\textsubscript{A1} = T\textsubscript{0} + \frac{d}{c+v} + \delta \textsubscript{B} - (T\textsubscript{0}+\delta\textsubscript{A})\]
which is equivalent to
\begin{equation} \label{eq1}
    T\textsubscript{B1}-T\textsubscript{A1} = \frac{d}{c+v} + (\delta \textsubscript{B} - \delta\textsubscript{A})
\end{equation}

device A also knows
\(T\textsubscript{A2}-T\textsubscript{B1},\) where \[T\textsubscript{A2} = T\textsubscript{2} + \delta \textsubscript{A} = T\textsubscript{1} +  \frac{d}{c-v} + \delta \textsubscript{A}\]
such that

\begin{equation} \label{eq2}
\begin{split}
    T\textsubscript{A2}-T\textsubscript{B1} & = T\textsubscript{1} + \frac{d}{c-v} + \delta \textsubscript{A} - 
    (T\textsubscript{1} + \delta \textsubscript{B}) \\ & = \frac{d}{c-v} + (\delta \textsubscript{A} - \delta \textsubscript{B})
\end{split}
\end{equation}

By adding \ref{eq1} and \ref{eq2} we obtain
\begin{equation} \label{eq3}
\begin{split}
T\textsubscript{A2}-T\textsubscript{A1} = \frac{d}{c+v} + \frac{d}{c-v} & = \frac{(dc-dv)+(dc+dv)}{c^2-v^2} \\ & = \frac{2dc}{c^2-v^2} = \frac{2d/c}{1-(v^2/c^2)}
\end{split}
\end{equation}

which, given $c$, gives us a good estimate of $d$ without any involvement of the clock drifts, however large. If $v$ is roughly 10\textsuperscript{-3}$c$, we get an estimate of $d$ with error on the order of 10\textsuperscript{-6}.



\end{appendices}


\bibliography{sn-bibliography}


\end{document}